# Study of muon tomographic imaging for high Z material detection with a Micromegas-based tracking system

Liu Cheng-Ming,[1] Wen Qun-Gang,[2, *] Zhang Zhi-Yong,[1, †] and Huang Guang-Shun [1, ‡]

**Affiliations:**

[1] State Key Laboratory of Particle Detection and Electronics, University of Science and Technology of China, Hefei 230026, China

[2] AnHui University, Hefei 230061, China

**Abstract:** A high spatial resolution tracking system was setup with the Micro-mesh gaseous structure (Micromegas) detectors in order to study the muon tomographic imaging technique. 6 layers of 90 mm × 90 mm one-dimensional readout Micromegas were used to construct a tracking system. The imaging test using some metallic bars was performed with cosmic ray muons. A two-dimensional imaging of the test object was presented with a newly proposed ratio algorithm.

## 1. Introduction

Muon imaging is a new technique developed to reconstruct images of volumes. According to the behaviors of muons in material, losing energy and absorbed in the material or changing direction after multiple Coulomb scattering, there are two different kinds of muon imaging, muon radiography and muon tomography imaging A well-defined characteristic of this technique using cosmic ray muons is its non-invasiveness and economical value [1][2]. Since its introduction in the 1950s [3], Muon imaging has been under development for many purposes, such as observation of volcanic activity [1], non-destructive exploration of historical sites [4], monitoring nuclear waste containers [2] and potential underground sites used for carbon sequestration [5], detecting cargo containers at custom entrances [6] etc.. This work is based on the muon tomography imaging.

Muons on Earth are the secondary particles that come from the extensive atmosphere shower of high-energy cosmic rays from space, with a flux about 1 cm$^{-2}$min$^{-1}$ at sea level and an average energy of about 3 GeV. Thus, high efficiency and good spatial resolution tracing detector is required for timely imaging events. For instance, a GEANT4-based simulation shows that better than 200 μm spatial resolution is required for the imaging of a liter-sized (10 cm x 10 cm x 10 cm) uranium cores shielded on each of their six sides by 2.5 cm of material with lower Z (Al or Pb) which were placed at different coordinates within a muon tomography station (3 m x 3 m x 5 m)[6].

A micro-mesh gaseous structure (Micromegas) is a typical Micro-pattern gas detector (MPGD), which has good spatial resolution of < 100 μm, is easy to extend to large area [7][8]. In this paper, a high spatial resolution tracking system consisting of 6 layers of one-dimensional readout Micromegas detectors is setup to study the muon imaging method for high-Z materials. The schematic design, detector and electronics, data acquisition (DAQ) of the test system is demonstrated. A new ratio algorithm is introduced and applied on the real data. A two-dimensional imaging of a test object is finally obtained.

## 2. Tracking system

*2.1. Schematic design*

In this tracking system, two 150 mm × 150 mm scintillators are used to select effective events that penetrate the whole tracking detectors and provide trigger signals for the DAQ by coinciding their signals. Six layers of 90 mm × 90 mm one-dimensional readout Micromegas detectors are used to determine the tracks of incident and exiting muons from a test object. As shown in Figure 1, the detectors 4-6 are for the tracking of incident muons; detectors 1-3 are for exiting muons scattered by the nuclei of the test object. Signals of the Micromegas detectors are readout by front-end electronics with APV25 ASICs [9] and digitized by MPD data acquisition system [10].

The test object consists of four layers of different metallic bars including copper, iron, aluminum and tungsten, with the atomic mass A of 64, 56, 27 and 183, respectively. Each layer has the same three metallic bars with a size of 10 mm × 10 mm × 100 mm. In other words, it is a 30 mm width, 40 mm high, and 100 mm long test object in total, and is placed along the readout strips of Micromegas detectors.

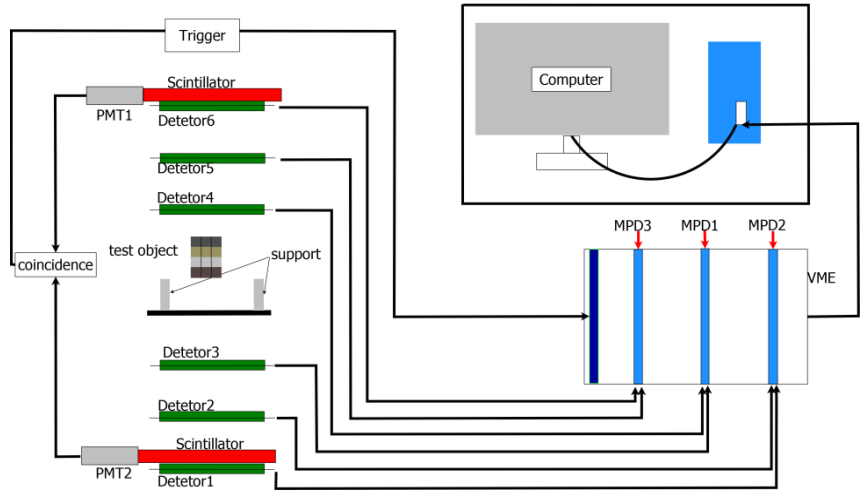

Figure 1. Schematic design diagram for muon imaging study

## 2.2. Setup of the system

The Micromegas detectors are manufactured using the thermal bonding method [11][12]. A resistive anode is employed to enhance the gas gain to higher than $10^4$. The readout of detectors is designed as one-dimensional strips with a pitch of 412 μm and drift region of 5 mm, the gas mixture is 93% argon to 7% carbon dioxide, As a consequence, better than 150 μm spatial resolution and higher than 90% detection efficiency are obtained for the cosmic ray muons. As shown in Figure 2, each Micromegas detector is installed in an aluminum plane, where a 100 mm × 100 mm hollow hole under the active area of detector is made to minimize the scattering materials in the tracker of muons.

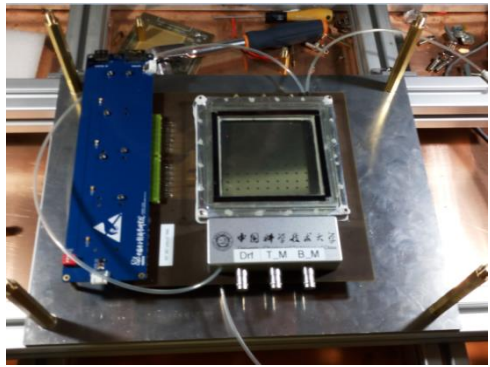

Figure 2. Installation of the tracking system for a single detector

The six detector planes are subsequently stacked to construct the tracking system as the schematic design mentioned above. Figure 3 is the finished view of the installation of Micromegas detectors, and the test object is inserted in the middle of the tracking system. Finally, the whole system is calibrated and aligned with cosmic test, and effective muon events are collected for imaging analysis.

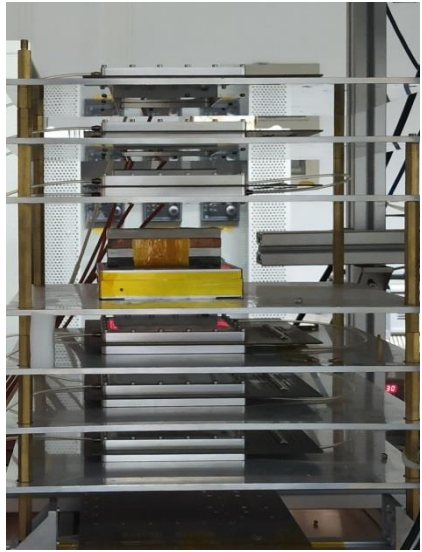

Figure 3. The finished view of installation of the tracker detectors

3. **Imaging with the cosmic ray muons**

*3.1.  Ratio algorithm for imaging*

In this experiment, a new ratio algorithm is proposed for imaging with the effective muon events, based on the idea of calculating the root mean square error value of the deflection angles of incoming and outgoing tracks of events in an exploration area. The deflection angle θ of muon passing through the material approximates a Gaussian distribution with a mean value of zero when considering the multi-Coulomb scattering as single scattering as Equation 1 shows. Equation 2 gives an expression of the root mean square error $\sigma_\theta$.

L is the length of a muon track in a certain material (air or test object). In this work, size of 33 mm x 33 mm is taken for the exploration area, considering the limited acceptance angle the L value could be treated as a constant.

$L_{rad}$ is the radiation length of the material, while the commonly used empirical formula is defined as Equation 3, where Z and A are the atomic number and atomic mass of the substance, respectively and β, c, p are the Lorentz parameters (β~1), the velocity of light, and the momentum of muons, respectively.

$$f(\theta) = \frac{1}{\sqrt{2\pi}\sigma_\theta} \exp(-\frac{\theta^2}{2\sigma_\theta^2}) \quad (1)$$

$$\sigma_\theta = \frac{13.6\text{MeV}}{\beta c p} \sqrt{\frac{L}{L_{rad}}} [1 + 0.038 \ln(\frac{L}{L_{rad}})] \quad (2)$$

$$L_{rad} = \frac{716.4 \cdot A}{Z(Z+1)\ln(287/\sqrt{Z})} [g/cm^3] \quad (3)$$

The variables in Equation 2 are momentum p and radiation length $L_{rad}$. In the case where momentum p is fixed value, as the atomic number Z increases the radiation length of the material decreases in Equation 3, and the $\sigma_\theta$ increases accordingly in Equation 2.

Defining H as Equation 4, Equation 2 can subsequently be written like Equation 5.

$$H = \frac{13.6\text{MeV}}{\beta c} \sqrt{\frac{L}{L_{rad}}} [1 + 0.038\ln(\frac{L}{L_{rad}})] \quad (4)$$

$$\sigma_\theta^2 = \frac{1}{p^2} H^2 \quad (5)$$

For a distribution with a mean value of zero, its root mean square error can be obtained by fitting the experiment data into Equation 6, where the $\theta_i$ represents the measured scattering angle of the number i of muon events in the data, and N the total number of events.

$$\sigma_\theta^2 = \frac{1}{N} \sum_{i=1}^{N} \theta_i^2 \quad (6)$$

Next, a test object which has a certain H value and whose N events can be divided into n groups according to the momentum of muons. For a group of muons in momentum bin $p_i$ (i=1, 2… n), supposing there are $k_i$ events with scattering angle $\theta_{i1}$, $\theta_{i2}$… $\theta_{ik_i}$. A series of the root mean square errors can be obtained through Equation 5 and 6:

$$\sigma_1^2 = \frac{1}{k_1} \sum_{j=1}^{k_1} \theta_{1j}^2 = \frac{1}{p_1^2} H^2$$

$$\sigma_2^2 = \frac{1}{k_2} \sum_{j=1}^{k_2} \theta_{2j}^2 = \frac{1}{p_2^2} H^2$$

$$\dots$$

$$\sigma_n^2 = \frac{1}{k_n} \sum_{j=1}^{k_n} \theta_{nj}^2 = \frac{1}{p_n^2} H^2$$

Rearranging these equations leads to Equation 7, where the coefficient $\frac{k_i}{N}$ indicates the probability of muons in momentum bin $p_i$.

$$\sigma_\theta^2 = \frac{1}{N}\sum_{i=1}^{N}\theta_i^2 = \sum_{i=1}^{n}\frac{k_i}{N}\frac{1}{p_i^2}H^2 \quad (7)$$

When the amount of experimental data is sufficient, the momentum distribution can be considered to satisfy the natural distribution at sea level [13], such that the term $\sum_{i=1}^{n}\frac{k_i}{N}\frac{1}{p_i^2}$ can be considered as a constant value A which is indeed the average value $<\frac{1}{p^2}>$ when N is large enough. Equation 7 consequently becomes:

$$\sigma_\theta^2 = \frac{1}{N}\sum_{i=1}^{N}\theta_i^2 = AH^2 \,.(8)$$

Compared to Equation 5, the influence to root mean square error from the momentum of muons is negligible.

An intuitive method for distinguishing different materials is to calculate the root mean square error of the muon scattering angles with Equation 8 and it names RMS error imaging method in this work. Due to the term which sums the square of the scattering angle, RMS error imaging method is sensitive to large angle events. In order to eliminate the effect of large angle and obtain good quality of imaging, it is necessary to constrain the angle range.

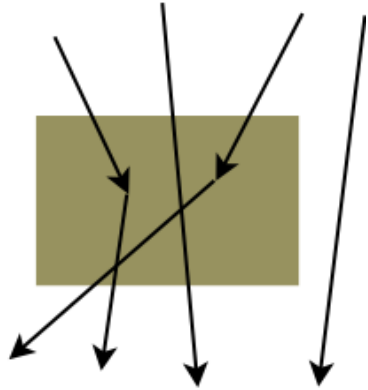

Figure 4. Muon event pass through the test object

To avoid the influence of large angle events, the so-called Ratio Algorithm was proposed. Figure 4 shows the behaviors of muons passing through the test object. As mentioned in this paper, the deflection angles obeys Gaussian distribution as Figure 5 shows, the red line indicating the angle distribution of muons passing through the low atomic number Z material, while the blue line relates to the high Z material. As all the events detected are counted as $A_0$, and events in certain angle range such as $|\theta|<\theta_0$ are counted as $A_c$. The ratio value R is defined as Equation 9

$$R = \frac{A_c}{A_0} \quad (9)$$

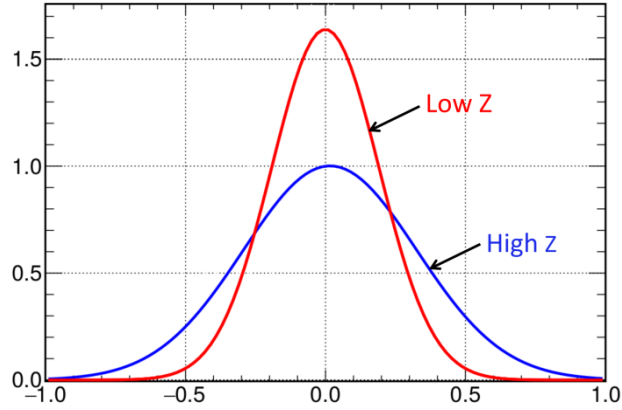

Figure 5. The Gaussian distribution of muon deflection angles in different materials

The ratio value is actually the Gaussian function definite integral (or likelihood error function) of the root mean square error $\sigma_\theta$, as shown in Equation 10:

$$R(\theta_0) = \frac{1}{\sqrt{2\pi}\sigma_\theta} \int_{-\theta_0}^{\theta_0} \exp\left(-\frac{\theta^2}{2\sigma_\theta^2}\right) d\theta \quad (10)$$

Obviously, for materials with different Z value, the higher Z material has smaller R value. In this way, different materials (air and test object in this work) can be distinguished. Furthermore, the R values could also be linked to the Z values if the tracks were precise enough.

### 3.2. *Evaluation with real data*

The ratio algorithm is proposed for eliminating the effects of large angle events. It is subsequently applied to experimental data. As shown in Figure 6, the schematic diagram of the test system with the test object and the positions along the y-axis are measurable, from top to bottom, the locations of six detectors are -3.5 mm, 58.9 mm, 106.5 mm, 266.8 mm, 320.8 mm, 383.8 mm. When muons are scattered by a test object passing by all of the six detectors, the position of the x-axis can be obtained through the detector readout. The incoming and outgoing tracks can subsequently be reconstructed and, as a result, the two-dimensional position of the exact scatter point is found. For imaging of the test object, ratios for all points in the measurement area are needed. As an example, the ratio at point A in Figure 6 is calculated through the

following steps:

- Take a square exploration area with a random point A as its center. Here the exploration area is 33 mm × 33 mm.
- All events passing through the exploration area are counted as $A_0$, and events with scatter angles in a certain range ($|\theta| < 1.6°$ in this work) are counted as $A_c$
- Use Equation 9 to get the ratio R.

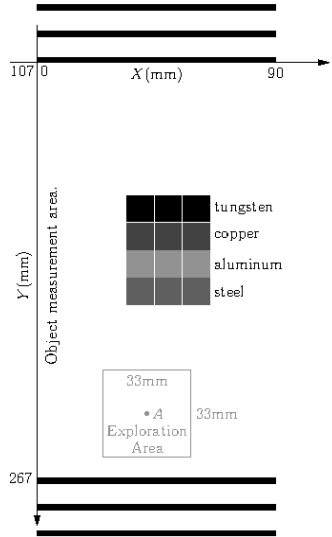

Figure 6. The Diagram of the description of the Ratio Algorithm

All of the events counted as $A_0$ that pass through the exploration area can be divided into two situations by studying the incoming and outgoing tracks. 1) for a straight track, whether it pass the top margin or the bottom margin will be recorded, 2) for a scattered track, if the scattering point is in the exploration area it will be recorded, if not, then it will be discarded. Then these scattered tracks satisfy the selection criteria ($|\theta| < 1.6°$) are counted as $A_c$. This shows that the ratio algorithm can make good use of these events which have scattering angle near 0°.

In order to ensure the unbiased R values, 500,000 points were selected randomly in the test object measurement area, and the R values were calculated for all points. Consequently, a quantity field R(x, y) (i.e. the R value with coordinates around point A) of the object measurement area is obtained. In this way, the image of cosmic-ray muon tomography is visualized using ratios, as shown in Figure 7. The rectangular box with coordinates shows the location and size of the test object. These points with ratio less than 0.800 are roughly formed into a rectangle. The reconstructed rectangle is smaller than the actual test object, but the reconstructed position is consistent with the actual position. Besides, these points with a ratio greater than 0.880 represent the air. The ratio value from 0.800 to 0.880 mainly occurs in the transition region between the test object (high-Z) and air (low-Z). In this system, the reconstruction uses data from 793 effective muons detected in one day, with a detection flux about 10 $cm^{-2}$ $day^{-1}$. The lower detection flux is due to removing the noised events caused by the detectors'' performance. Improving detection flux will be very useful for rapid detection to some extent.

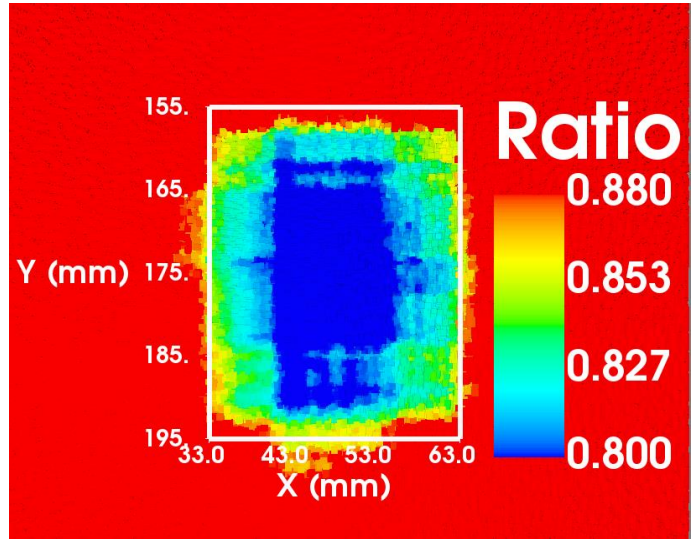

Figure 7. The image of test object by cosmic ray muon tomography

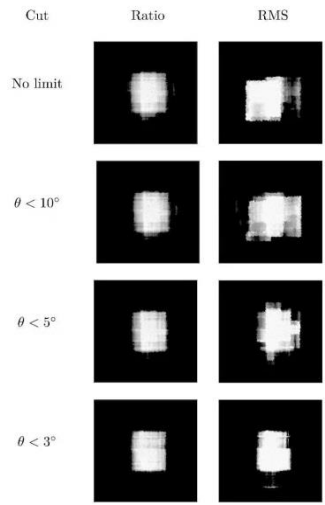

Figure 8. Comparison of Ratio and RMS error imaging method after the different angle limits

As mentioned above, constraining the scattered angle to suppress the effect of large angles is one feature of RMS error imaging method. These two algorithms are applied under different scatter angle θ cuts for eliminating the large angle events, as shown in Figure 8. It can be seen that more stringent angle cuts bring better imaging quality for RMS error imaging method, but less influence to ratio algorithm.

**Discussion**

The muon tomography imaging is a very promising technique for non-invasiveness detecting with many potential applications. This work realized imaging with a tracking system based on Micromegas detectors and a new proposed imaging algorithm named ratio algorithm which can make good use of events with scattering angles near 0°.

Development of this technique is ongoing. The focus of our work is on identifying different high Z materials with ratio algorithm and on improving the performance of tracking system to achieve large acceptance and to get good quality data. In particular, 2 dimension readout detectors is under developing for new tracking system in order to achieve 3 dimension muon tomography imaging.

**Acknowledgements**

This work was supported by the Program of National Natural Science Foundation of China Grant No. 11605197, the Fundamental Research Funds for the Central Universities, and the State Key Laboratory of Particle Detection and Electronics, SKLPDE-ZZ-201818, SKLPDE-KF-201912. This work was partially carried out at the USTC Center for Micro and Nanoscale Research and Fabrication, and we wish to thank Yu Wei for his help on the nanofabrication steps for germanium coating.